\def\ref{\par\noindent\hang}

\def\spose#1{\hbox to 0pt{#1\hss}}
\def\approxlt{\mathrel{\spose{\lower 3pt\hbox{$\sim$}}
        \raise 2.0pt\hbox{$$<$$}}}
\def\approxgt{\mathrel{\spose{\lower 3pt\hbox{$\sim$}}
        \raise 2.0pt\hbox{$>$}}}

\def\multleft#1{\hbox to size{\vbox {\halign {\lft{##}\cr #1}}\hfill}\par}
\def\multright#1{\hbox to size{\vbox {\halign {\rt{##}\cr #1}}\hfill}\par}

\def\today{\ifcase\month\or January\or February\or March\or April\or May\or
      June\or July\or August\or September\or October\or November\or December\fi
      \space\number\day, \number\year}
\def\$<${\thinspace}
\def\s{\hbox{\phantom{5}}}      

\def\boxit#1{\vbox{\hrule\hbox{\vrule\kern3pt\vbox{\kern3pt
          #1 \kern3pt}\kern3pt\vrule}\hrule}}

\def\cm{{\rm\thinspace cm}}

\def\erg{{\rm\thinspace erg}}
\def\eV{{\rm\thinspace eV}}

\def\ph{{\rm\thinspace ph}}
\def\s{{\rm\thinspace s}}

\def\ergpcmsqps{\hbox{$\erg\cm^{-2}\s^{-1}\,$}}

\def\phpcmsqps{\hbox{$\ph\cm^{-2}\s^{-1}\,$}}

\documentstyle[psfig,times]{mn}
\begin{document}
\hsize=6truein

\def\simless{\mathbin{\lower 3pt\hbox
   {$\rlap{\raise 5pt\hbox{$\char'074$}}\mathchar"7218$}}}   
\def\simgreat{\mathbin{\lower 3pt\hbox
   {$\rlap{\raise 5pt\hbox{$\char'076$}}\mathchar"7218$}}}   
\def\anisotropy{\frac{\Delta T}{T}}

\title[ASCA and RXTE Long Look at MCG--6-30-15]{First Constraints on Iron Abundance versus Reflection Fraction from the Seyfert~1 Galaxy MCG--6-30-15 }

\author[]
{\parbox[]{6.in} { J.C.~Lee,$^1$ A.C.~Fabian,$^1$ W.N.~Brandt,$^2$
C.S.~Reynolds$^3$ and K.~Iwasawa$^1$ \\
\footnotesize \it $^1$ Institute of Astronomy; Madingley Road; Cambridge CB3 0HA \\
\it $^2$ Department of Astronomy and Astrophysics; The Pennsylvania State University; 525 Davey Lab; University Park, PA 16802 USA \\
\it $^3$ JILA; Campus Box 440; University of Colorado; Boulder, 80309-0440 USA  \\
}}

\maketitle

\begin{abstract}  
We report on a joint ASCA and RXTE observation spanning an $\sim$ 400~ks
time  interval of the bright Seyfert~1 galaxy MCG--6-30-15.  The data
clearly confirm  the presence of a broad skewed iron line
($W_{K\alpha} \sim$ 266~eV) and Compton reflection continuum at higher
energies reported in our previous paper.  We also investigate whether the
gravitational and Doppler effects that affect the iron line may also
be manifest in the reflected continuum. 
The uniqueness of this data set is underlined by the extremely good
statistics that we obtain from the approximately four million photons
that make up the 2-20~keV   RXTE PCA spectrum alone.  This, coupled
with the high energy coverage of HEXTE and the spectral resolution of ASCA in
the iron line regime has allowed us to constrain the relationship
between abundance and reflection fraction for the first time at the 99
per cent confidence level.  The reflection fraction is entirely
consistent with a flat disk, i.e.  the cold material subtends $\rm 2
\pi$ sr at the source, to an accuracy of 20 per cent.  Monte Carlo
simulations show that the observed strong iron line intensity is
explained  by an overabundance of iron by a factor of $\sim$ 2 and an
underabundance of the lower-Z elements by a similar factor.  By
considering non-standard abundances, a clear and consistent picture
can be made in which both the iron line and reflection continuum come
from the same material such as e.g. an accretion disk.

\end{abstract}


\section{INTRODUCTION} 
The current paradigm for active galactic nuclei (AGN) is a central
engine consisting of an accretion disk surrounding a supermassive
black hole (e.g. see review by Rees 1984).  The main source of energy
is gravitational potential energy as material falls in and is heated
to high temperatures in some sort of dissipative disk.

The accretion disk is assumed to consist of {\it cold} optically thick
material.  {\it Cold} in this context means that iron is more neutral
than Fe XVII (oxygen is not fully ionized although H and He may be
ionized).  Depending on the geometry, this material may be subjected
to irradiation.   Careful study of the X-ray reprocessing mechanisms
can give much information about the immediate environment of the
accreting black hole.  These effects of reprocessing can often be
observed in the form of emission and absorption features in the X-ray
spectra of AGNs.  In Seyfert 1 nuclei, approximately half of the
X-rays are `reflected' off the inner regions of the accretion disk
(Guilbert \& Rees 1988; Lightman \& White 1988), and superposed on the
direct (power-law) primary X-ray emission.  The general consensus is
that the power-law component is emitted in the corona above  the disk.
As photons pass through the corona, some fraction will be upscattered
to X-ray energies.  Multiple Compton scatterings tend to produce a
power-law X-ray spectrum.  The principle observables of the reflection
spectrum are a fluorescent iron K$\alpha$ line, and Compton
backscattered continuum which hardens the observed spectrum above
$\sim$ 10~keV.

\begin{figure*}
\psfig{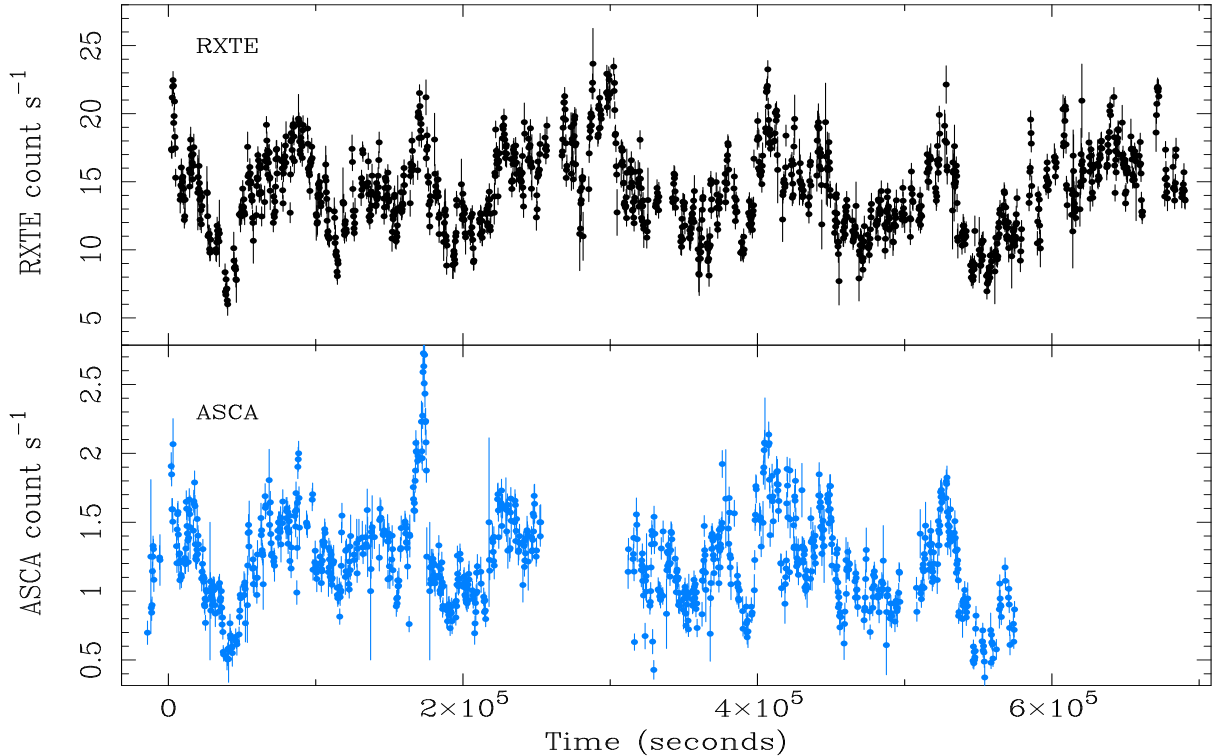}
\caption[h]{Background-subtracted light curve of MCG$-$6-30-15 for observations in the RXTE PCA 2--60~keV band, and the ASCA SIS 0.5-10~keV band. The epoch of the start and stop times for RXTE and ASCA respectively is 1997 August 4 to 1997 August 12, and 1997 August3 to 1997 August 10. }
\end{figure*}

The iron line together with the reflection component are important
diagnostics of the X-ray continuum source. The strength of the emission
line relative to the reflection hump between 20--30~keV is largely
dependent on the abundance of iron relative to hydrogen in the disk
(George \& Fabian 1991; hereafter GF91), as well as the normalization
of the reflection spectrum relative to the direct spectrum.  The
relative normalization of the reflection spectrum probably depends
primarily on the geometry (i.e. the solid angle subtended by the
reflection parts of the disk as seen by the X-ray source).  However,
strong light bending effects (e.g. Martocchia
\& Matt 1996) or special-relativistic beaming effects
(e.g. Reynolds \& Fabian 1997, Beloborodov 1999) can also enhance the amount of
reflection.

MCG--6-30-15 is a bright nearby ($z=0.0078$) Seyfert 1 galaxy that has
been extensively studied by every major X-ray observatory since its
identification. An extended {\it EXOSAT} observation provided the
first evidence for fluorescent iron line emission (Nandra et al. 1989)
which was attributed to X-ray reflection.  Confirmation of these iron
features by {\it Ginga} as well as the discovery of the associated
Compton reflected continuum supported the reflection picture (Nandra,
Pounds \& Stewart 1990; Pounds et al. 1990; Matsuoka et al. 1990)
while ASCA data have shown the iron line to be broad, skewed, and
variable (eg. Tanaka et al 1995; Iwasawa et al. 1997). The high energy
and broad-band coverage afforded by  the {\it Rossi X-ray Timing
Explorer (RXTE)}  show convincingly the simultaneous presence of both
the broad iron line and reflection component for the first time in
this object (Lee et al. 1998).  Data on MCG--6-30-15 from BeppoSAX
also confirm the presence of a broad skewed iron line and reflection
continuum (Guainazzi et al. 1998).
 
We present in this paper a long look at MCG--6-30-15 simultaneously by
RXTE and ASCA spanning a time interval of $\rm \sim 400~ks$. (The RXTE
on-source time was $\sim$ 400~ks and ASCA on-source time was  $\rm
\sim 200~ks$).  Our observations clearly  confirm the presence of a
redshifted broad iron K$\alpha$ line at $\sim$ 6.3~keV and reflection
hump between 20--30~keV.  The high energy instrument HEXTE on RXTE
coupled with ASCA's sensitivity at the lower energies allow us to
constrain for the  first time the relationship between reflection
fraction and abundance values at the 99 per cent confidence level.  We
explore these features of reflection and iron emission in detail.
An in-depth investigation of variability is beyond the scope of this
paper and will be addressed in a later paper.

\section{Observations}
MCG--6-30-15 was observed by RXTE spanning $\sim$ 400~ks over the
period from 4 Aug 1997 to 12 Aug 1997 by both the Proportional Counter
Array (PCA) and High-Energy X-ray Timing Experiment (HEXTE)
instruments. (The final useable integration times were 304~ks for the
PCA and  114~ks for HEXTE.)  It was simultaneously observed for $\rm
\sim 200~ks$ (with useable intergration time  of 197~ks) by the ASCA
Solid-state Imaging  Spectrometers (SIS) over the period 1997 August 3
to 1997 August 10 with a half-day gap part-way through the observation
(PI : H. Inoue).  The SIS was operated in Faint mode throughout the
observation, using the standard CCD chips (S0C1 and S1C3).    We
concentrate primarily on the RXTE spectra in this paper.

\subsection{RXTE and ASCA Instruments}
The Rossi X-Ray Timing Explorer (RXTE) consists of three instruments.
The two pointed instruments are the Proportional Counter Array (PCA) that 
covers the lower energy range and the High Energy X-ray Timing Experiment
(HEXTE) that covers the higher energies.  Together, the two instruments cover
the energy band between 2 and 200~keV.
The PCA consists of 5 Xenon Proportional Counter Units (PCUs)
sensitive to X-ray energies between 2--60~keV with $\sim 18 $ per cent
energy resolution at 6~keV.  The total collecting area is 6500 $\rm
cm^2$  ($\sim$ 3900 $\rm cm^2$ for 3 PCUs) with a $1^\circ$ FWHM field
of view.   The HEXTE instrument is coaligned with the PCA and covers
an energy range between 20--200~keV.  For a more thorough review
on these instruments, we refer the reader to Jahoda et al. (1996) and
Rothschild et al. (1998).

The Advanced Satellite for Cosmology and Astrophysics (ASCA) is a 
Japanese X-ray observatory that was launched on 1993 February 20,
designed and constructed as a joint endeavor with the United States.
It consists of four identical grazing-incidence X-ray telescopes,
each terminating with a fixed detector.  The focal plane detectors are 
two CCD cameras (Solid state Imaging Spectrometer, or SIS) and two
gas scintillation imaging proportional counters (Gas Imaging Spectrometer,
or GIS). All four detectors are operated simultaneously all the time.
The ASCA SIS is sensitive in the energy range between 0.4 and
10~keV, with an energy resolution of 2 per cent at 5.9~keV. Its  field
of view is 22 $\rm arcmin^2$.  The primary goal of the SIS is spectroscopy
in the 0.4-10~keV energy band. Its PSF is completely determined by the 
telescope rather than the detector response.  For a more in-depth discussion,
we defer to Tanaka, Inoue \& Holt (1994).

\subsection{Data Analysis}
\subsubsection{RXTE Reduction}

We extract PCA light curves and spectra from only the top Xenon layer
using the newly released Ftools 4.1 software.  Data from PCUs 0, 1,
and 2 are combined to improve signal-to-noise at the expense of
slightly blurring the spectral  resolution.  Data from the remaining
PCUs (PCU 3 and 4) are excluded because these instruments are known to
periodically suffer discharge and are hence  sometimes turned off.

Good time intervals were selected to exclude any earth or South
Atlantic Anomaly (SAA) passage occulations, and to ensure stable
pointing.  We also filter out electron contamination events.

We generate background data using {\sc pcabackest v2.0c} in order  to
estimate the internal background caused by interactions between the
radiation/particles and the detector/spacecraft at the time of
observation.  This is done by matching the conditions of observations
with those in various model files.  The model files used are the L7-240
 background models which are intended to be specialized for application
to faint sources with count rate less than 100 cts/sec. 

The PCA response matrix for the RXTE data set was created using {\sc
pcarsp  v2.36}.  In the course of performing the spectral fitting
described in this section, we discovered a bug in the pcarmf package.
This resulted in the 1998-Aug-29 memo
(http://legacy.gsfc.nasa.gov/docs/xte/whatsnew/calibration.html\#c
als41b)  from NASA-GSFC detailing the circumstances under which pcarmf
does not properly account for the temporal changes in the response
matrices.  All spectral fitting presented here has been corrected for
this software bug.  Background models and response matrices are
representative of the most up-to-date PCA calibrations.

The net HEXTE spectra were generated by subtracting spectra of the
off-source positions from the on-source data.  Time intervals were
chosen to exclude 32 seconds prior to and 320 seconds following SAA
passages.  This conservative approach avoids periods when the internal
background is changing rapidly.  Data in which the satellite elevation
is less than 10 degrees above the Earth's limb is also excluded.  
We use the standard 1997 March 20 HEXTE response matrices provided by 
the RXTE Guest Observer Facility (GOF) at Goddard Space Flight Center.
The relative normalizations of the PCA and the two HEXTE clusters are 
allowed to vary, due to uncertainties ( $<$ about 5\% ) in the HEXTE 
deadtime measurement.

\subsubsection{ASCA Reduction}

ASCA data reduction was carried out using {\sc FTOOLS} version 4.0 and
4.1 with standard calibration provided by the ASCA GOF.  Detected SIS
events with a grade of 0, 2, 3 or 4 are used for the analysis.  One of
the standard data selection criteria, {\sc br earth}, elevation angle
of the source from the bright Earth rim, does not materially affect
the hard ASCA data while it does contribute to the soft X-ray data
from the SIS at some level.
 We use the SIS data of
approximately 231 ks from each detector for spectral analysis.  The
source counts are collected from a region centred at the X-ray peak
within $\approx$ 4 arcmin from the SIS and 5 arcmin from the GIS.  The
background data are taken from a (nearly) source-free region in the
same detector with the same observing time.

\begin{table}
\begin{center}
\begin{tabular}{|c|c|c|c|}
\multicolumn{4}{c}{\sc Detection Statistics} \\
\hline
{\em \rm Instrument} & {\em $\rm ^{a} Energy Band$} & {\em $\rm
^{b} Count Rate$} & {\em $\rm ^{c} Flux$} \\
\hline
\hline
ASCA S0 / S1 & 0.6-10 & 1.16 / 0.93 & -  \\
   "  & 0.5-2 & -  & 1.47  \\
   " & 2-10 & - & 3.41  \\
RXTE PCA & 2-10 & - & 4.92  \\
RXTE PCA & 3-20 & 12.72 & 6.26  \\
HEXTE ClsA / ClsB & 16-200 & 0.73 / 0.63 &   \\
HEXTE ClsA / ClsB & 16-50 & - & 2.7 / 2.9  \\
\hline

\end{tabular}
\caption{ $^a$ Energy Band in units of keV. 
$^b$ Average count rate for given energy band in units of ct/s.
$^c$ Flux in given energy band in units of $10^{-11}$ \ergpcmsqps}

\end{center}
\end{table}

\begin{figure}
\psfig{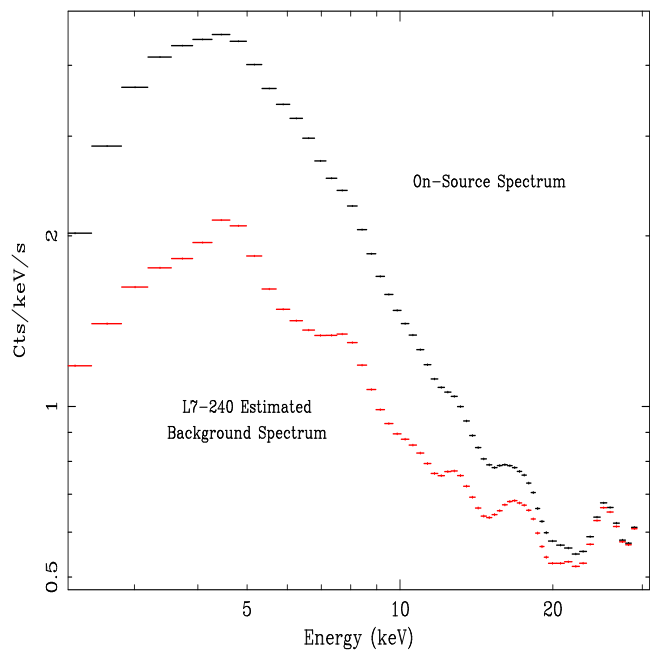}
\caption[h]{Plot of on-source and estimated background spectra in the energy range 2 keV $<$ E $<$ 30 keV.  The background model used
is L7-240 appropriate for faint sources and high voltage epoch3.}
\end{figure}

Table~1 details the average count rate and fluxes for specified
energy bands as detected by the ASCA S0 and S1 instruments, and the
RXTE PCA and HEXTE Cluster A and Cluster B detectors.
 
Fig.~1 shows the ASCA S0 160-2700 pha-channel ($\approx$
0.6--10~keV),  and the RXTE PCA 1--129 pha channel ($\approx$
2-60~keV) background subtracted light curves.   There is a gap of
$\sim$ 60~ks in the ASCA light curve in which the satellite observed
IC4329A while MCG--6-30-15 underwent a large flare observed by RXTE.
Significant variability can be seen in both light curves on short and
long timescales, which will be investigated in detail in a later
paper.  Flare and minima events are seen to correlate temporally in
both light curves.

\section{Spectral Fitting}
We fit the data in a number of ways in order to investigate the known
features of fluorescent iron emission (eg. Fabian et al. 1994, Tanaka
et al. 1995) and Compton reflection (eg. Lee et al. 1998;
Pounds et. al. 1990, Matsuoka et al. 1990; Nandra
\& Pounds 1994) in this object.


\begin{figure}
\psfig{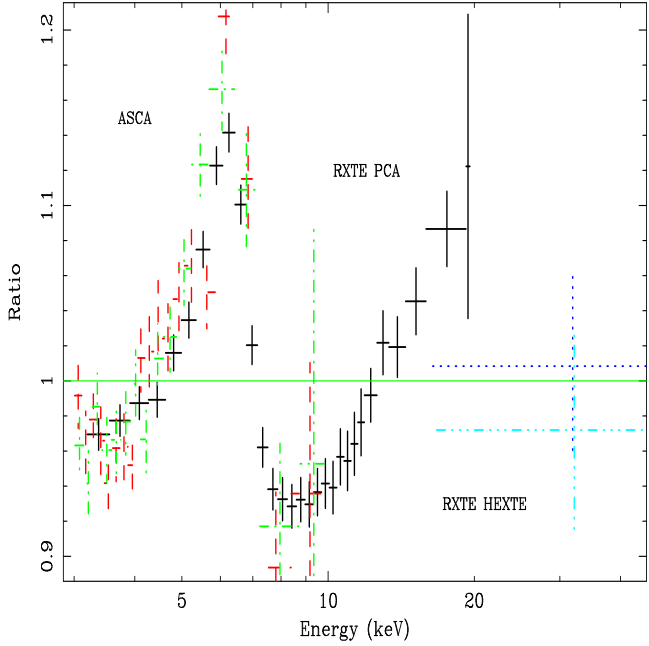}
\caption[h]{Ratio of data to continuum using a nominal simple power-law fit confirms the clear existence of a redshifted broad iron line at $\sim$ 6.3~keV and reflection component above 10~keV.  The simultaneous RXTE and ASCA fit demonstrates the good agreement between the two satellites in the  energy band below 10~keV.}
\end{figure}

\begin{figure}
\psfig{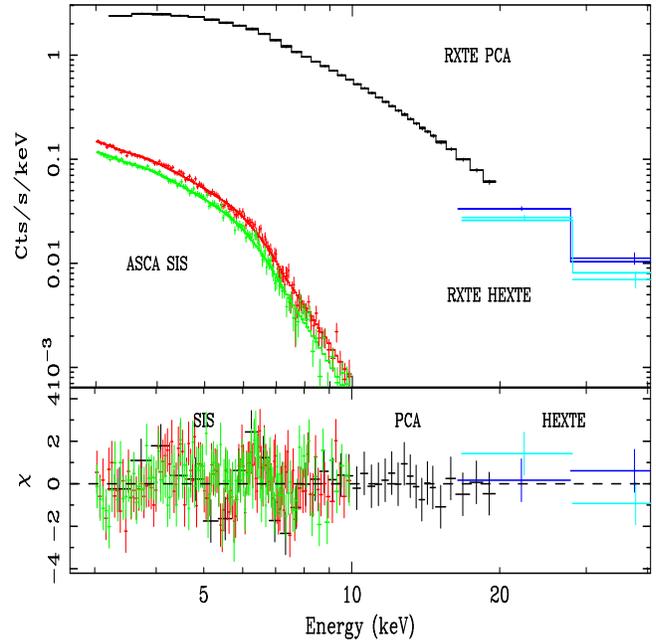}
\caption[h]{Joint ASCA/PCA/HEXTE spectrum of the simultaneous data from
MCG--6-30-15 provide further evidence for the existence of the reflection component and good agreement between
 ASCA and RXTE. A multicomponent model fit to all three data sets that include a Gaussian component to model the iron line and a reflected spectrum shows that the residuals are essentially flat.}
\end{figure}

We restrict ASCA and PCA data analysis to be respectively between 3
and 10~keV, and 3 and 20~keV (Fig.~2 shows that the PCA background
dominates above 25~keV).   This lower energy restriction at 3~keV is
selected in order that the necessity for modeling photoelectric
absorption due to Galactic ISM material, or the warm absorber that is
known to be present in this object is removed.  This also allows us to
bypass recent problems with residual dark currents in the ASCA
0.5--0.8~keV energy band, and RXTE calibration problems that may still
exist at energies below 2~keV.  Despite the fact that  these
absorption features are only important below  $\sim$ 2~keV, we
nevertheless check their significance in two ways : (1) we fix the column 
density at the value of $\rm 4.09$ $\rm
\times$ $\rm 10^{20}$ $\rm {cm}^{-2}$ to account for Galactic ISM
absorption along the line of sight for this object, and (2) we allow this 
parameter to be free.  For both cases, we find that the difference 
in best fit values between a model that includes and one that excludes
this absorption effect is negligible.  Additionally, for the latter case
we find that the model is unable to place any 
tight constraints on the column density in the chosen energy 
range for the RXTE data.
Accordingly, we neglect this parameter from our fits.
We have also checked that the standard background-subtraction
methods described in Section 2.2.1 should adequately account for the
PCA background in the energies of interest.  As added checks for the 
quality of the reduction and background subtraction, we extract 
spectra from 81~ks of Earth-occulted data ({\sc elv $\le$ 0}), and find
for the occulation data that the normalized flux per keV is zero for the
selected energy range. 
HEXTE data are restricted to be between 16 and 40~keV
in order that we may adequately model the reflection hump.  HEXTE
response matrices are inadequate below 16~keV  (William Heindl, 1998
private communication); 0.5 per cent  systematics were added to PCA
spectra.

\subsection{Spectral Features}
A nominal fit to all three data sets (i.e. ASCA, RXTE PCA and HEXTE)
demonstrates the clear existence of a redshifted broad iron K$\alpha$
line at $\sim$ 6.3~keV and reflection hump between 20 and 30~keV as
shown in Lee et al. 1998.  Fig.~3 further demonstrates the good
agreement between ASCA and RXTE at energies below 10~keV, and in
particular, at the energies surrounding the iron line.

As added assurance for the existence of the reflection component and
good agreement between ASCA and RXTE, a multicomponent model fit that
includes the reflected spectrum to all three data sets shows that the
residuals are essentially flat (fig.~4); a gaussian component is used to 
account for the iron line.  The underlying continuum is
fit with the model {\sc pexrav} which is a power law with an
exponential cut off at high energies reflected by an optically thick
slab of neutral material \cite{mz}.  We fix the inclination angle of
the reflector at $30^\circ$ so as to agree with the disk inclination
one obtains when fitting accretion disk models to the iron line
profile as seen by {\it ASCA} (Tanaka et al. 1995). The high energy
cutoff is fixed at 100~keV consistent with thermal corona models.  We
perform fits in which the high energy cutoff is allowed to be a free
parameter and find that RXTE is incapable of placing any contraints on
this value for MCG--6-30-15 (the best fit values can vary anywhere
from $\sim$ 30~keV to a few hundred keV). For this reason, we rely on
the values determined by BeppoSAX. (The Phoswitch Detector System PDS
on board BeppoSAX has a better spectral sensitivity than the RXTE
HEXTE at the higher energies.)  For robustness, we further test the
sensitivity of RXTE to the high energy cutoff by performing fits in
which this parameter is allowed to vary within  the 100--400~keV 90\%
confidence region as determined by BeppoSAX (Guainazzi et al. 1998).
The preferred high energy cutoff is 100~keV with infinity as the upper
limit error.  We also test for the significance of the high energy
cutoff value in the determination of fit parameters; this is done by
comparing fit results for 100~keV and 400~keV  cutoff energies ; the
two results do not differ with any statistical significance.  We fit
the RXTE data in the 3--40~keV energy range with a multiple  component
model consisting of a power-law reflection and redshifted  Gaussian
component.  Errors are quoted at the 90 per cent confidence  ($\Delta
\chi^2$ = 2.71, Bevington \& Robinson 1994).

The best fit values for the 100~keV cutoff energy case are detailed in
Table~2.  (A double gaussian fit using the ASCA parameters shows a
negligible improvement of $\Delta \chi^2 $ = 3 for one extra
parameter.)  For comparison, the best fit values for which this cutoff
energy was fixed at 400~keV are : power-law slope $\Gamma$ =
$2.07^{+0.06}_{-0.04}$ with a power law flux at 1~keV $A =
(1.95^{+0.13}_{-0.10}) \times 10^{-2}$ $\rm ph$ $\rm cm^{-2}$ $\rm
s^{-1}$. The reflection fraction is $0.81^{+0.16}_{-0.15}$ for lower
elemental abundances set equal to that of iron of
$0.72^{+0.26}_{-0.17}$ solar abundances.  The redshifted line energy
is  $\rm 5.98 \pm 0.08~keV$ with line width $\sigma$ = $\rm
0.58^{+0.11}_{-0.10}~keV$ and intensity of the iron line I =
$1.62^{+0.26}_{-0.19} \times 10^{-4}$ \phpcmsqps.   The equivalent
width $W_{K\alpha}$ = $\rm 300^{+48}_{-35}~eV$ and $\chi^2$ is  39 for
47 degrees of freedom.  While the fit values between the two cutoff
energy scenarios differ by little, it is clear that a high energy
cutoff of 100~keV is preferred ($\Delta \chi^2 \sim 4$ for no additional
parameters).  Accordingly, the results that follow
are based on this fixed value for  the high energy cutoff.  We note
that the $\chi^2$ values are likely to be lower than expected due to
the tendency of the reduction software to overestimate the errors.


\begin{figure}
\psfig{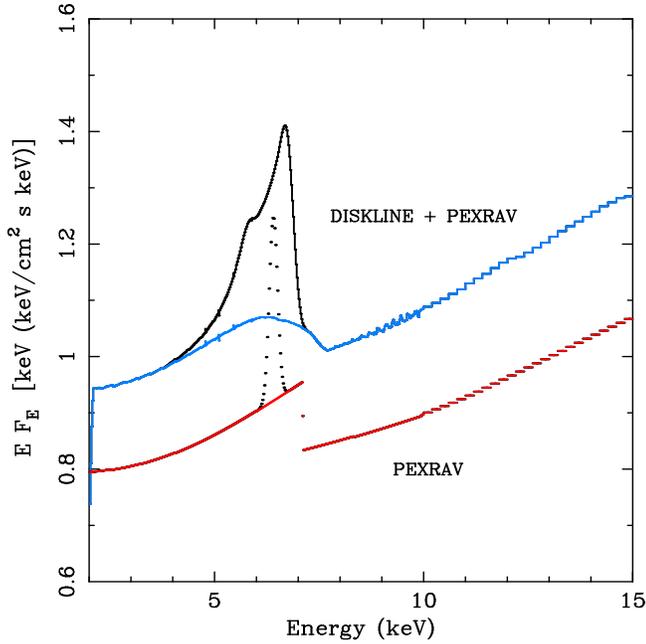}
\caption[h]{A model that accounts for relativistic smearing of the reflection component with disk-line model parameters (upper plot) as contrasted with a
model that does not include Doppler and gravitational effects on the
reflected component (lower plot).  Note the contrast in the sharpness
of the iron edge at $\sim$ 7~keV.  The two models have been
artificially renormalized for the sake of illustration.}
\end{figure}


\subsection{Doppler and Gravitational effects on the Reflected Spectrum}

We next investigate the degree to which gravitational and Doppler
effects which determine the line profile also affect the reflection
continuum shape (the previous section does not account for this
effect).  We use a convolution model ({\sc rdblur}) that assumes the same
characteristics as the disk-line model for a Schwarzschild geometry by
Fabian et al. (1989) for application to the reflected spectrum.  (The
reflected continuum is convolved with the same kernel as the diskline
model.)  For these fits, we assume a cold accretion disk inclined at
$i=30^\circ$  (Tanaka et al. 1995).  The radial emissivity, $\alpha$
assuming a power-law-type emissivity function ($\propto R^{- \alpha}$)
of the line, is left as a free parameter.  The innermost radius of
stable orbit $R_{in}$ for a
Schwarzschild geometry is set to Iwasawa et al. (1999)
best fit ASCA value $6.7 r_g$ ; the outermost radius $R_{out}$ is 
left as a free parameter ($r_g  \equiv GM/c^2$ is
the gravitational radius of the black hole), for this object.
A comparison between the differences in the two models is shown in fig.~5.

We test for the effects of gravity and Doppler shifts by comparing the 
quality of the fits using (1) a model that accounts for the effects of 
relativistically smeared reflection and (2) a model that does not 
account for this effect. For the latter, the {\sc diskline} model is used 
for the iron emission and {\sc pexrav} for the continuum : {\sc pexrav} + 
{\sc diskline}.  A similar model is used for the former with the addition
of a multiplicative component ({\sc rdblur}) that convolves the continuum with the same kernal 
as the diskline model  : {\sc rdblur}$\ast$({\sc pexrav})+{\sc diskline}.
We have investigated fits in which the lower-Z and iron abundances were tied together
and left as free parameters, and fits in which the former was fixed at 0.5 solar abundances
and the latter fixed at twice solar abundances appropriate for the $W_{K\alpha}$ value seen in
this object. (A more detailed investigation of the effects of abundances on the $W_{K\alpha}$ and 
reflection fraction is presented in subsequent sections.)
The results for the latter (fixed abundances) 
vacillate between statistically insignificant preferences for relativistically
smeared reflection from RXTE data ($\Delta \chi^2 = 1.5$ corresponding to 1 extra `hidden' parameter,
for 49 degrees of freedom), and the contrary from ASCA data
($\Delta \chi^2 = 1.5$ for 681 degrees of freedom). We have also
searched for this effect in the 1994 ASCA data and find that $\Delta \chi^2 = 0.2$ for 681 degrees
of freedom. When the low-Z and iron abundances are untied and left as free parameters, 
the RXTE data prefer smeared reflection with $\Delta \chi^2 = 2.7$ for 47 degrees of freedom. Additionally,
while derivations for the iron abundance remains at twice solar, there is a preference for slightly higher
low-Z abundances for the case of smeared reflection.
$\Delta \chi^2$ values from ASCA remain simliar to those previously mentioned when abundances were 
fixed; there is a clear indication that ASCA is insensitive to abundance measurements.
The reflection fraction is fixed at unity. 

While we suspect that relativistic effects are indeed prevalent in the reflected
spectrum, it appears that the sensitivity of the present data are insufficient 
for detecting the effects of relativistically smeared reflection with large
statistical significance.  We suspect that any deviation in $\chi^2$ is largely due to the
the modeling of the iron edge at $\rm \sim 7\,keV$ as illustrated in
Fig.~5.   The  contrast is remarkably noticeable at that energy
between the models :  the  standard {\sc pexrav} model that we use for
our fits invokes a  sharp edge at those energies which is neither seen
in the time-averaged ASCA nor RXTE data.  The sharpness and depth of
this feature is diminished when we invoke Doppler and gravitational
effects. However, we are trying to detect a 10 per cent effect which is also at the level
of the ASCA error bars above $\sim$ 7~keV. Differences in $\chi^2$ can come
additionally from the smearing of the reflection hump, and may be partly the reason
that RXTE with its higher energy coverage prefers smeared reflection. 

We have checked our findings 
against the  level of present calibration uncertainties
in the PCA response matrix.  An absorbed power-law and narrow Gaussian
fit to a 47 ks archived data of the quasar 3C~273 from the same gain epoch
(epoch 3) as our observations give residuals less than 1 per cent for
the energies of interest.
Furthermore, the best fit 3C~273 values are comparable to those obtained with ASCA and 
BeppoSAX observations of this object (e.g. Haardt et al. 1998; 
Orr et al. 1998; Cappi et al. 1998).  
However, because fit results from a model that includes the effects of 
relativistically smeared reflection and one that does not differ by
little, we also present results for the former whenever appropriate 
in this paper.  Due to the insensitivity of RXTE  to the line
profile, the results in which relativistically smeared reflection is considered,
presented in Table~2 and subsequently correspond to the model :
{\sc rdblur}$\ast$({\sc pexrav + narrow gaussian}) with $\alpha$, $R_{in}$ and
$R_{out}$ frozen respectively at 2, $6 r_g$, and $30 r_g$  The best fit results 
are similar to those derived from the {\sc diskline} models mentioned
previously. 

\subsection{Constraints on Reflection Fraction and Iron Abundance}
Having established that the reflection component exists, we next
investigate the  relationship between the iron abundance and
reflection fraction for the reflection scenarios described in sections
3.1 and 3.2.  In order to obtain a better understanding of physical
models of AGN central regions, we need to disentangle the abundance
from the absolute normalization of the reflection component.  Due to
the lack of  good broad spectral coverage in previous observations,
the fit parameters of  reflection fraction, elemental abundances, and
power-law index were strongly coupled.  With the high energy coverage
of HEXTE and ASCA's good spectral resolution at energies between 0.6
and 10~keV, we can decouple these parameters for the first time and
study the relationship each has with respect to the others.  Fig.~6
shows the confidence contours for abundance versus reflection fraction
as expected  from a corona+disk model.  The best fit value for
abundance and reflection fraction for the standard {\sc pexrav} case
are respectively $0.92^{+0.31}_{-0.21}$ solar abundances and
$1.09^{+0.26}_{-0.19}$; the corresponding values for smeared
reflection are $1.13^{+0.31}_{-0.21}$ solar abundances and
$1.16^{+0.40}_{-0.26}$.  Both results are consistent with the scenario
in which the primary X-ray source is  above the accretion disk
subtending an angle of 2$\pi$ sr  (i.e. corresponding to a reflection
fraction of $\frac{\Omega}{2 \pi} = 1$).

\begin{figure}
\psfig{file=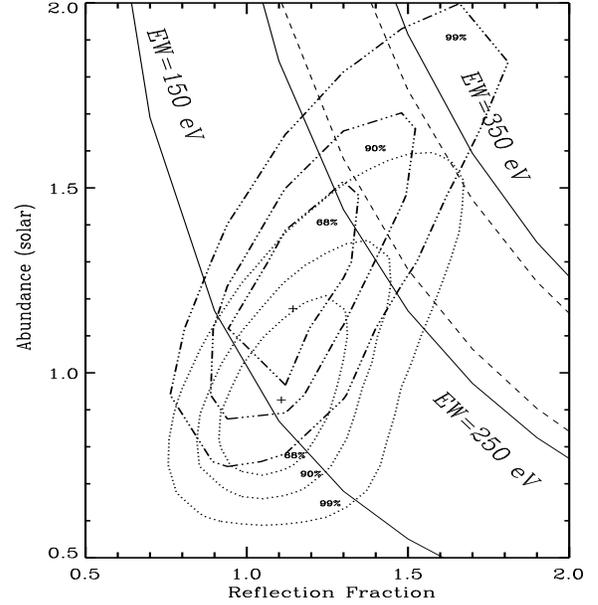,angle=0,width=8.5truecm,height=8.5truecm}
\caption[h]{Fitting in the energy range 3~keV $\le$ E $\le$ 40~keV, we constrain upper bounds
for the abundance vs. reflection fraction relationship at the 99 per cent confidence level for the first time.  The dot-dashed-line contours correspond to fits in which relativistically smeared reflection (due to Doppler and gravitational effects) is considered; the dotted-line contours do not include this effect. Solid lines represent contours of constant equivalent width; dashed lines correspond to the equivalent width values of $266^{+46}_{-33}$~eV and $331 \pm 25$~eV respectively for non-smeared and relativistically smeared reflection.} 
\end{figure}

We wish to stress the uniqueness of this data set for both its good
statistics and energy coverage.  In the RXTE PCA 2-20~keV spectrum
alone, we estimate that $\approx$ 4 million photons make up the
spectrum for our 400~ks exposure.  (The 2--20~keV flux is $\sim$ 8.6
photons $\rm cm^{-2}\, s^{-1}$ for a detector effective  area at those
energies of $\rm \sim 1000\, cm^2$, for 3 PCUs.)  Its uniqueness is
underlined by our ability to place combined constraints on both the
line and reflection fraction.  However, it should be noted that {\sc
pexrav} does not model the iron  emission feature which, for the case
of RXTE data is achieved using a  Gaussian line profile.  Accordingly,
the consistency of the strength of the line with {\sc pexrav}
predictions for the resulting absorption is  investigated via Monte
Carlo simulations in the next section.  GF91 have shown that the
strength of the iron K-shell absorption features can be increased by
varying the elemental abundances, or  by ionizing the lower atomic
weight elements.  The enhanced iron abundance causes more iron K-shell
absorption of the reflection continuum thereby weakening the
reflection continuum as the abundance rises.

For completeness, we also consider the significance of the inclination
angle value for our determination of the reflection fraction and
abundance results. (We test this only for the non-smeared reflection
case since the determination of the inclination is largely due to the
iron line peak rather than the reflected continuum.) This test is also
relevant for assessing the accuracy of the value we obtain for
$W_{K\alpha}$, which is dependent on the inclination and $\Gamma$;
ASCA's determination of the inclination from this observation is
$31^\circ \pm 2^\circ$ (Iwasawa et al. 1999, submitted) consistent
with previous observations (Tanaka et al. 1995).  For exaggerated
inclinations of 20$^\circ$ and 40$^\circ$, we obtain respectively
reflection fraction and abundance values to be $1.02^{+0.25}_{-0.18}$
and $0.91^{+0.32}_{-0.21}$, and $1.18^{+0.26}_{-0.21}$ and
$0.95^{+0.33}_{-0.22}$ which are within the errors of the values
obtained from fits in which we fix the inclination at 30$^\circ$.
The apparent insensitivity of RXTE to the inclination value is due to
the inability of the satellite to resolve the blue peak of the iron
line.  The inclination is determined at  the energy where the blue
peak of the line abruptly drops off; the spectral resolution of RXTE
is inadequate for resolving this feature in much detail. $\Gamma$
remains  relatively unchanged.

\begin{table*}
\begin{center}
\begin{tabular}{|c|c|c|c|c|c|c|c|c|c|c|c|}
\multicolumn{12}{c}{\sc Summary of Spectral Fits for Sections 3.1 -- 3.4}\\
\hline
{\em \rm Section} & {\em $\rm \Gamma_{4-20}^a$} & {\em $\rm refl^b$ } & {\em $\rm LowZ^c$ } & 
{\em $\rm Iron^d$ } & {\em $\rm A^e$ } & {\em $\rm LineE^f$ } & {\em $\rm \sigma^g$} & 
{\em $\rm I_{K \alpha}^h$} & {\em $\rm W_{K\alpha}^i$ } & {\em $\rm \chi^{2}/dof^j$} & {\em $\rm grav^{k}$} \\ 
\hline
\hline

3.1 & $2.02^{+0.05}_{-0.04}$ & $1.09^{+0.26}_{-0.19}$ & $0.92^{+0.31}_{-0.21}$ &
= abund & $1.88^{+0.11}_{-0.09}$ & $5.99^{+0.07}_{-0.09}$ & $0.54^{+0.11}_{-0.10}$ &
$1.45^{+0.25}_{-0.18}$ & $266^{+46}_{-33}$ & 35/47\\ 

3.2 & $ 1.99 \pm 0.04$ & $1.13^{+0.31}_{-0.21}$ & $1.16^{+0.40}_{-0.26}$ &
= abund & $1.91^{+0.11}_{-0.09}$ & $6.28 \pm 0.05 $ & $0.05^{+0.17}_{-0.05}$&
$1.74^{+0.13}_{-0.13}$ & $331^{+25}_{-25}$ & 24/47 & $\surd$ \\ 



3.4 & $2.01^{+0.04}_{-0.05}$ & {\bf 1} & $0.44^{+0.32}_{-0.11}$ &
$2.05^{+2.05}_{-0.99}$ & $1.80^{+0.10}_{-0.10}$ & $5.98 \pm 0.08$ & ${\bf 0.54}$ &
$1.14^{+0.29}_{-0.33}$ & $201^{+51}_{-58}$ & 31/49 \\ 

3.4 & $1.98 \pm 0.04$ & {\bf 1} & $0.82^{+1.03}_{-0.38}$ &
$1.50^{+2.05}_{-1.04}$ & $1.88^{+0.11}_{-0.11}$ & $  6.29^{+0.63}_{-0.57} $ & ${\bf 0}$ &
$1.66^{+0.30}_{-0.29}$ & $312^{+56}_{-55}$ & 24/49 & $\surd$ \\ 


\hline

\end{tabular}
\caption{Summary of fits as discussed in sections 3.1--3.4.  Bold-faced values indicate parameters that were fixed 
at the value given.  For all fit values indicated here, the redshift
and high energy cutoff were consistently fixed at the values of 0.0078
and 100~keV respectively, with inclination fix at best fit ASCA value
of $30^{\circ}$.  $^a$ Power-law photon index. $^b$ Reflective
fraction = $\Omega / 2\pi$. $^c$ Lower elemental abundances relative
to solar abundances. $^d$ Iron abundances relative to solar. $^e$
Power-law flux at 1 keV, in units of $10^{-2}$ $\rm ph $ $\rm cm^2 $
$\rm s^{-1}$ $\rm keV^{-1}$. $^f$ Energy of the iron $K \alpha$. $^g$
Iron line width (keV) $^h$ Intensity of iron emission line in units of
$10^{-4}$ \phpcmsqps.  $^i$ Equivalent width of the emission line in
units of eV. $^j$ $\chi^2$ per degree of freedom. $^k$ Column is
checked if relativistically smeared reflection is accounted for in the
fit. }

\end{center}
\end{table*}






\subsection{Iron Abundance and Strength of the Reflected Continuum}
Up until this point, we have treated the iron emission and absorption
(which has direct bearing on the derived reflection fraction) as
separate additive components of a multi-component model.  This is due
largely to {\sc pexrav} modelling only the reflection continuum, which
is imprinted with the absorption feature.  However, we need to assess
the consistency of the line intensity with {\sc pexrav} model
predictions of this absorption.  This is now discussed in the context
of Monte Carlo simulations.

Previous workers (i.e. GF91; Matt, Perola \& Piro 1991; Reynolds,
Fabian, \& Inoue 1995) have investigated the effect of abundance
values on the equivalent width of the line and the associated
reflected spectrum.  This was done via Monte Carlo simulations in
which incident photons are assigned a random initial energy with a
power-law distribution function $\Gamma \sim 1.9$, and an incident
energy (corresponding to an isotropic source).  The probabilities for
a photon to be either Compton scattered or photoelectrically absorbed
for a given energy are tracked. 

The equivalent width of the iron line in MCG--6-30-15 is $\rm
266^{+46}_{-33}~eV$ and $\rm 331 \pm 25~eV$ for non-smeared and
smeared reflection respectively, when the reflector has $\sim$ solar
abundance and reflection fraction close to unity.  Both are
inconsistent with the predicted $W_{K\alpha}$ of $\rm 150~eV$ from
GF91 for a slab of cold material subtending $\rm 2 \pi~sr$ at the
X-ray source, rotating around a static black hole. We show in Fig.~6
the locus of points in the abundance -- reflection plane which has an
equivalent width of $250\pm100\eV$. It is clear that the highest
reflection fraction part of the diagram leads to a much stronger line
than is observed. The best-fitting solution lies within the region
defined by the reflection-fitting contours and the lines of observed
equivalent width. 

Some fraction of an iron line can be due to reflection from outer
material (the putative torus say). This fraction is however likely to
be small in MCG--6-30-15 since the part of the line consistent with a
zero-velocity narrow core sometimes disappears at some phases of the
ASCA data (Iwasawa et al 1996).

The equivalent width of the line does not depend however on the iron
abundance alone, but also upon the abundance of the lower-Z elements
such as oxygen which can absorb the line before it emerges from the
reflector. We have therefore investigated the behaviour of the
solution in the multi-dimensional, iron abundance, lower-Z abundance,
reflection fraction space. To make the problem tractable we adopt unit
reflection fraction. This is indeed the preferred solution when the
iron and lower-Z abundances are separated. The contours in the
abundance plane for the (non-smeared) results are shown in Fig.~7,
showing that separate abundances are indeed preferred.  We then use a
Monte-Carlo code to track the predicted equivalent width along the
major locus of the abundance contours. The equivalent widths are shown
in Fig.~8 and reveal that a value of 300~eV occurs when the iron
abundance is about twice the solar value and the lower-Z abundances
are about half the solar value.  Details of actual fit values using
the {\sc pexrav} and relativistically smeared reflection models can be
found in Table~2.

\begin{figure}
\psfig{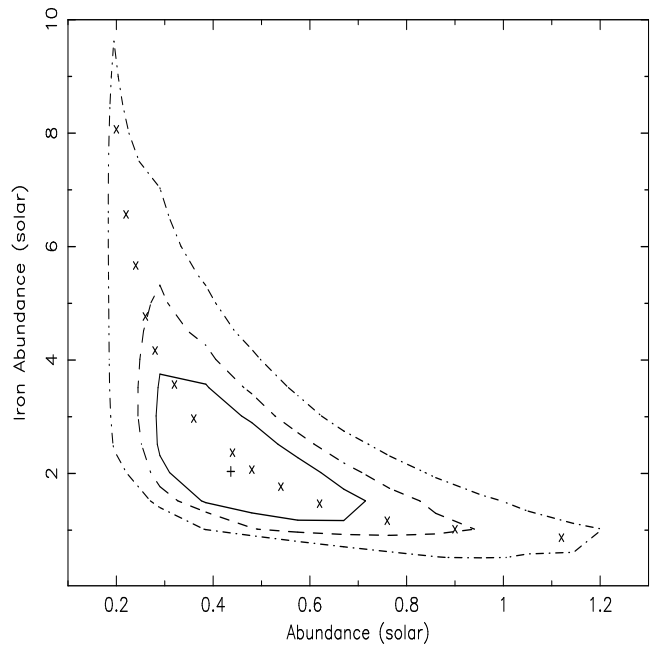}
\caption[h]{Confidence contours (corresponding to 68, 90, and 99 per cent confidence levels) for iron abundance versus lower elemental abundances for material subtending 2$\pi$ sr at the X-ray source. The '+' symbol correspond to the best fit values ; the 'x' symbols are used for illustrative purposes for Fig.~8.}
\end{figure}

\begin{figure}
\psfig{file=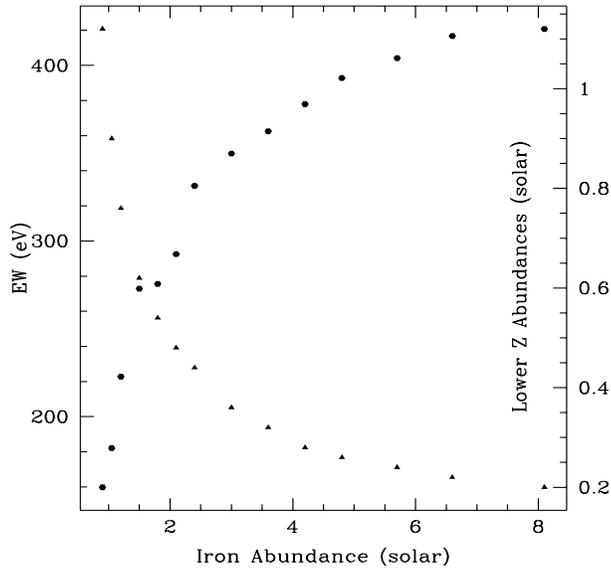,angle=0,width=9.0truecm,height=8.0truecm}
\caption[h]{Filled hexagons correspond to plot of equivalent width of the fluorescent iron K$\alpha$ line
as a function of iron abundance for 30$^\circ$ inclination.  The
results are based on Monte Carlo simulations in which the low-Z
abundances and iron abundance values (as defined by Morrison \&
McCammon 1983) are modified according to the locus of points
illustrated by the symbol 'x' in the contour plot of Fig.~7. The
chosen abundance and iron abundance  values used in the simulations
are represented by filled triangles.}
\end{figure}

\section{Discussion}
This simultaneous long observation coupled with the combined strengths
of ASCA and RXTE have enabled us to constrain for the first time the
relationship between iron abundance and reflection fraction at the 99
per cent confidence level, as well as confirm the presence of a broad
skewed iron line for MCG--6-30-15.  With the additional high energy
coverage of HEXTE, we establish that the features of reflection are
present and that results are consistent with the scenario in which
cold material subtends $2 \pi$ sr at the X-ray source. We further
investigate the effects of gravity and Doppler shifts on the reflection
component, but find that both RXTE and ASCA are insensitive to this.
Additionally, we verify that the effects of the cutoff energy  do not
compromise our results. By fitting the data with the 100~keV lower
limit for the cutoff energy and comparing that to the 400~keV cutoff
energy fit results, we find that they are consistent with each other
within their errors.  The preferred cutoff energy however is 100~keV.

Monte Carlo simulations further reveal that an overabundance of iron
by a factor of $\sim$ 2 is needed to reconcile the large value
for the equivalent width that we observe for both the standard and
relativistically smeared reflection scenarios; the equivalent width is
even more dramatically enhanced when relativistivistic effects are
invoked.  By considering non-standard abundances, a consistent picture
can be made for which both the iron line and reflection continuum
originate from the same material / structure such as, e.g. an
accretion disk.  We find also that the factor of two to three iron
overabundance as predicted by our data holds consistently even in
comparisons with models of FeII emission, known to strongly contribute
to the optical and UV continuum of many active objects (Wills, Netzer,
\& Wills 1985). It is also note-worthy to consider the importance of
abundance determinations for assessing the chemical history of the
host galaxy.  For example, an iron rich environment with depleted
amounts of lighter elements (as suggested in our data) may provide
evidence that Type Ia supernovae events were likely to have occurred
in high proportions during the history of the galaxy.  

\section{ACKNOWLEDGEMENTS}
We thank all the members of the RXTE GOF for answering our inquiries
in such a timely manner, with special thanks to William Heindl and the
HEXTE team for help with HEXTE data reduction.  We also thank Keith
Jahoda for explanations of PCA calibration issues, and Roderick Johnstone
and Keith Arnaud for their time and help with software.   JCL thanks the
Isaac Newton Trust, the Overseas Research Studentship programme (ORS)
and the Cambridge Commonwealth Trust for support.  ACF thanks the
Royal Society for support.
WNB thanks the NASA RXTE grant NAG5-6852, and the NASA Long Term Space Astrophysics 
(LTSA) grant NAG5-8107 for support.
CSR thanks the National  Science Foundation
for support under grant AST9529175, and NASA for support under the
LTSA grant NASA-NAG-6337.  KI thanks PPARC.

\end{document}